\documentclass[a4paper,fleqn,usenatbib]{mnras}

\usepackage{newtxtext,newtxmath}
\usepackage[T1]{fontenc}
\usepackage{ae,aecompl}

\usepackage{natbib}
\usepackage{amssymb}

\usepackage{graphicx}

\newcommand{\msun}{\ensuremath{\, \mathrm{M}_{\sun{}} }}
\newcommand{\zsun}{\ensuremath{\, \mathrm{Z}_{\sun{}} }}

\newcommand{\mzams}{\ensuremath{ M_{\mathrm{ZAMS}}}}
\newcommand{\mfin}{\ensuremath{ M_{\mathrm{fin}}}}
\newcommand{\mhef}{\ensuremath{ M_{\mathrm{He,f}}}}
\newcommand{\mcof}{\ensuremath{ M_{\mathrm{CO,f}}}}
\newcommand{\mrem}{\ensuremath{ M_{\mathrm{rem}}}}
\newcommand{\ap}{\ensuremath{\alpha_{\mathrm{P}}}}
\newcommand{\sevn}{\textsc{SEVN}}
\newcommand{\simA}{set--A}
\newcommand{\simB}{set--B}
\newcommand{\parsec}{\textsc{PARSEC}}
\newcommand{\starlab}{\textsc{Starlab}}
\newcommand{\higpus}{\textsc{HiGPUs}}
\newcommand{\higpusr}{\textsc{HiGPUs-R}}

\title[Very massive stars, PISNe and IMBHs]{Very massive stars, pair-instability supernovae and intermediate-mass black holes with the \sevn{} code}

\author[Spera \& Mapelli]{
	Mario Spera,$^{1,2}$\thanks{E-mail: mario.spera@oapd.inaf.it (MS)} 
	Michela Mapelli$^{1,2}$\\
	$^{1}$INAF, Osservatorio Astronomico di Padova, Vicolo dell'Osservatorio 5, I-35122, Padova, Italy\\
	$^{2}$INFN, Milano Bicocca, Piazza della Scienza 3, I-20126, Milano, Italy\\}

\date{Accepted XXX. Received YYY; in original form ZZZ}

\begin{document}
	
	\label{firstpage}
	\pagerange{\pageref{firstpage}--\pageref{lastpage}}
	\maketitle
	
	\begin{abstract}
		Understanding the link between massive ($\gtrsim 30\msun{}$) stellar black holes (BHs) and their progenitor stars is a crucial step to interpret observations of gravitational-wave events. In this paper, we discuss the final fate of very massive stars (VMSs), with zero-age main sequence (ZAMS) mass $>150$ M$_\odot$, accounting for pulsational pair-instability supernovae (PPISNe) and for pair-instability supernovae (PISNe). We describe an updated version of our population synthesis code \sevn{}, in which we added stellar evolution tracks for VMSs with ZAMS mass up to $350\msun{}$ and we included analytical prescriptions for PPISNe and PISNe. We use the new version of \sevn{} to study the BH mass spectrum at different metallicity $Z$, ranging from $Z=2.0\times 10^{-4}$ to $Z=2.0\times 10^{-2}$. The main effect of PPISNe and PISNe is to favour the formation of BHs in the mass range of the first gravitational-wave event (GW150914), while they prevent the formation of remnants with mass 60--120 $\msun{}$. In particular, we find that PPISNe significantly enhance mass loss of metal-poor ($Z\leq 2.0\times 10^{-3}$) stars with ZAMS mass $60\leq \mzams{}/\msun{}\leq 125$. In contrast, PISNe become effective only for moderately metal-poor ($Z<8.0\times 10^{-3}$) VMSs. VMSs with $m_{\rm ZAMS}\gtrsim{}220$ M$_\odot$ and $Z<10^{-3}$ do not undergo PISNe and form intermediate-mass BHs (IMBHs, with mass $\gtrsim 200\msun{}$) via direct collapse. 
\end{abstract}
	

	\section{Introduction}

All three gravitational wave (GW) events detected so far by the Advanced Laser Interferometer Gravitational-Wave Observatory (aLIGO) have been interpreted as the merger of a black hole (BH) binary \citep{abbott2016a, abbott2016c, abbott2017}. Additionally, aLIGO identified another BH merger candidate (LVT151012) even though its significance is below the threshold to claim an unambiguous detection \citep{abbott2016d}. The aLIGO detections demonstrated that stellar BH binaries (BHBs) exist and can merge within a Hubble time \citep{abbott2016b}. Besides GW150914, GW151226 and GW170104, the confirmed stellar BHs are only few tens, most of them observed in Milky Way's X-ray binaries \citep{ozel2010}. Dynamical mass measurements in X-ray binaries, possible  only for about a dozen of BHs, suggest a dearth of stellar BHs with mass $\gtrsim 15\msun{}$ \citep{farr2011,casares2014}. In contrast, the masses of the two BHs in GW150914 (GW170104) are $36.2^{+5.2}_{-3.8}\msun{}$ and $29.1^{+3.7}_{-4.4}\msun{}$ ($31.2^{+8.4}_{-6.0}\msun{}$ and $19.4^{+5.3}_{-5.9}\msun{}$), respectively, and their merger formed a new BH with mass $62.3^{+3.7}_{-3.1} \msun{}$ ($48.7^{+5.7}_{-4.6}\msun{}$).

Inferring the properties of the progenitors of such massive BHs is still an open issue. The mass of a compact object is expected to strongly depend on the evolution of its progenitor and on the final supernova (SN) mechanism \citep{mapelli2009,belczynski2010,mapelli2013,spera2015}. In the last decade, the models of stellar winds underwent a major upgrade \citep{vink2001,graefener2008,vink2011,muijres2012,vink2016}, which radically changed the landscape of massive star evolution (e.g. \citealt{paxton2013,paxton2015,chen2015}). In particular, mass loss by stellar winds solely determines the pre-SN mass (\mfin{}) of a star \citep{woosley2002}, which is a crucial ingredient to understand its final fate.

While the physical mechanisms powering core-collapse SNe are still matter of debate (see \citealt{janka2012} for a review), a dearth of observations of progenitor stars with zero-age main sequence (ZAMS) mass $\mzams{} \gtrsim 18\msun{}$ suggests that stars with a higher $\mzams$ might end their life quietly, without SN explosion (see \citealt{smartt2015} and references therein). Several theoretical models (e.g. \citealt{fryer1999,fryer2001,mapelli2009,fryer2012}) predict the possibility that a star collapses directly to a BH if its final mass is large enough ($\gtrsim{}30-40$ M$_\odot{}$, \citealt{spera2015}). Alternative models suggest that the possibility of a direct collapse depends on the innermost structure of a star at the onset of core-collapse \citep{oconnor2011,ertl2016}. Regardless of the discrepancies between SN models, a direct collapse seems to be the only viable scenario to explain the masses of GW150914 with stellar BHs.

Most models of the BH mass spectrum include the effect of core-collapse SNe but neglect the impact of pair-instability SNe (PISNe, \citealt{fowler1964}) and pulsational PISNe (PPISNe), with few remarkable exceptions \citep{woosley2017,belczynski2016}. Unlike core-collapse SNe, the physical mechanism powering PISNe \citep{ober1983,bond1984,heger2003,woosley2007} and PPISNe \citep{barkat1967,woosley2007,chen2014,yoshida2016} is quite well understood: if the mass of the Helium core ($\mhef{}$) is $\gtrsim 30\msun{}$, the formation of electron-positron pairs makes oxygen/silicon burn explosively.
Recent hydrodynamical simulations \citep{woosley2017} show that if $32\msun{}\lesssim{}\mhef{} \lesssim 64\msun{}$ the star undergoes several pulses that significantly enhance mass loss before the star forms a compact remnant (PPISN), while if $64\msun{}\lesssim{}\mhef{} \lesssim 135\msun{}$ the ignition of oxygen and silicon releases enough energy to disrupt the entire star (PISN).  If $\mhef{} \gtrsim 135\msun{}$, the star is expected to  avoid the PISN and to directly collapse.

Recently, \citet{woosley2017} investigated the onset of PPISNe for a grid of metal-poor stars $\left(Z\simeq 0.1\zsun{}\right)$ and for Helium-only stars. \citet{woosley2017} (hereafter W17) shows that PPISNe are expected to occur for $70\lesssim \mzams{}/\msun{} \lesssim 150$ while PISNe are effective for very massive stars (VMSs, $150 \lesssim \mzams{}/\msun{} \lesssim 260$). Stars with $\mzams{}\gtrsim 260\msun{}$ are expected to avoid the PISN and possibly form intermediate-mass BHs (IMBHs).

The dependence of these limits on the adopted stellar evolution prescriptions is unclear. The major uncertainties come from the evolutionary models of VMSs  ($\mzams{}\gtrsim 150\msun{}$, see e.g. \citealt{heger2003}). VMSs have been observed in extreme star forming regions (e.g. \citealt{crowther2016}) and might be the product of runaway collisions (i.e. multiple collisions in very dense stellar systems, \citealt{bonnell1998,portegieszwart1999}). VMSs are promising candidates to explode as PISNe \citep{kozyreva2017}. For instance, the luminosity curve of SN2007bi, SN 2213-1745 and SN 1000+0216 can be explained with a PISN model, assuming a progenitor star with a bare Helium core mass of $\sim 130\msun{}$, and a ZAMS mass of $\sim 250\msun{}$\citep{galyam2009,cooke2012}. Furthermore, VMSs have also been claimed to be the progenitors of IMBHs (\citealt{portegieszwart2002,portegieszwart2004,freitag2006}), but mass loss by stellar winds seems to be a key ingredient to understand their fate \citep{giersz2015,mapelli2016}. \citet{vink2011}  used a Monte Carlo approach to model the mass-loss rate of VMSs, but most stellar evolution codes do not include the \citet{vink2011} prescriptions, with the remarkable exception of the Padova And tRieste Stellar Evolution Code (\parsec{}, \citealt{chen2015}). As a consequence, the mass spectrum of heavy stellar BHs has been poorly investigated so far.

In this paper, we present an updated version of the  Stellar EVolution for N-body (\sevn{}) population-synthesis code. The new version of \sevn{} includes (i) an analytic treatment for PPISNe and PISNe, and (ii) up-to-date evolutionary tracks for VMSs ($\mzams{}$ up to $350\msun{}$, from \citealt{chen2015}). We use the new version of \sevn{} to study the BH mass spectrum for different metallicities and for $\mzams{}$ up to $350\msun{}$. We also discuss the impact of PPISNe and PISNe on the formation of the BH binaries observed by aLIGO and on the possibility of forming IMBHs from VMSs.

\section{method}\label{sec:method}
We updated \sevn{}  \citep{spera2015} to investigate the effect of PPISNe and PISNe on the BH mass spectrum. \sevn{} reads pre-evolved stellar evolution tracks, generated for a grid of ZAMS masses and different metallicity to calculate the physical properties of stars. The stellar tracks are given in the form of input tables that \sevn{} interpolates on--the--fly. This approach makes \sevn{} versatile because it is possible to change stellar evolution prescriptions by substituting the input tables, without modifying the internal structure of the code. 

In the last version of \sevn{} we have introduced a new, non-linear method to interpolate stellar evolution tracks. The new scheme significantly improves the old, linear algorithm since it allows us to use less points in the input tables and, at the same time, it reduces the interpolation error. The details of the new interpolation method are shown in Appendix \ref{app:B}.

The default version of \sevn{} includes a set of input tables generated using the \parsec{} code \citep{bressan2012,tang2014,chen2015}. The input tables range from metallicity $Z=2.0\times 10^{-4}$ to $Z=2.0\times 10^{-2}$. and each of them includes the evolutionary tracks of stars in the mass range $0.1\leq \mzams{}/\msun{} \leq 350$, with a mass step of $0.5\msun{}$ (the upper mass limit was $150$ M$_\odot$ in the previous version of \sevn{}, \citealt{spera2015}).

\sevn{} includes several up-to-date SN explosion prescriptions. Three of them (delayed, rapid and startrack models) are taken from \citet{fryer2012} and use the final (pre-SN) Carbon-Oxygen core mass of the star (\mcof{}) to distinguish between successful and failed SNe. The other two models are based on the compactness of the progenitor star at the onset of  core collapse \citep{oconnor2011,ertl2016}. The main difference between CO-based and compactness-based criteria is that the former have a net threshold (in terms of \mzams{}) between successful SNe and direct collapse, while the latter give a more complex picture (`islands of direct collapse') since the compactness of the progenitor star does not vary monotonically with the pre-SN mass of the star. 
 
Throughout this paper, if not specified otherwise, we adopt the delayed SN model from \citet{fryer2012}. This model predicts a successful SN, with fallback, for $\mcof{}<11\msun{}$ and direct collapse for $\mcof{}\geq 11\msun{}$. To distinguish between NSs and BHs we  fix an upper limit for the maximum NS mass: we assume that all remnants with mass $\mrem{} \geq 3 \msun{}$ are BHs \citep{oppenheimer1939,chamel2013}. When a NS or a BH forms, we also take into account the mass-loss due to the emission of neutrinos. For NSs, we follow the approach of \citet{timmes1996} who estimate that the mass lost in neutrinos is at the level of $\sim 0.1M_{\mathrm{bar}}$, where $M_{\mathrm{bar}}$ is the baryonic mass of the proto-compact object (see their eq. 8). For BHs, we follow \citet{fryer2012} who assume that BHs behave approximately the same way as NSs, so that the mass lost in neutrinos is simply fixed to $0.1M_{\mathrm{bar}}$.

To account for PISNe and PPISNe, we adopt the following approach. Since \sevn{} is not an hydrodynamical code, we cannot simulate PPISNe and PISNe  self--consistently. Still, we can use an analytical prescription that, starting from the physical properties of a star at a given time, mimics the mass loss due to PPISNe and PISNe.

The prescriptions implemented in \sevn{} are based on the recent results obtained by W17. We obtained a formula that fits the values of the compact remnant masses given in Tab.~1 and Tab.~2 of W17 with a relative error $\lesssim 5\%$. Tab.~1 of W17 shows the outcome of the explosion of Helium-only stars and Tab.~2 is for ordinary stars at low metallicity ($Z\simeq 0.1\zsun{}$). In Tab.~2, different mass loss prescriptions have been adopted to mimic the results for lower metallicity. Tab.~1 includes Helium stars with mass between $30\msun{}$ and $64\msun{}$. Tab. 2 shows the evolution of progenitor stars with $70\leq \mzams{}/\msun{}\leq 150$, corresponding to $30\lesssim \mhef{}/\msun{} \lesssim 70$. In absence of PPISNe and PISNe all these stars would have collapsed directly to massive BHs, losing only a small fraction of their final mass because of neutrino emission ($\sim 0.1\mfin{}$). 

In \sevn{} we adopt a fitting formula for the parameter 
\begin{equation}\label{eq:ap}
\ap{} \equiv \frac{\mrem{}}{M_{\mathrm{rem, no\,psn}}},
\end{equation}
where $M_{\mathrm{rem, no\,psn}}$ is the mass of the compact remnant we would obtain without PPISNe and PISNe. We express \ap{} as a function of the final Helium mass fraction of the star ($\mathcal{F} \equiv \dfrac{\mhef{}}{\mfin{}}$) and $\mhef{}$. We obtain the mass of the compact remnant as $\mrem{}=\ap{}M_{\mathrm{rem, no\,psn}}$, from Eq. \ref{eq:ap}. In particular, $\ap{}=1$ for remnants that form via direct collapse and $\ap{}=0$ for PISNe. The complete fitting formula is given  in Appendix \ref{app:A}.

The updated version of \sevn{}, including PISNe and PPISNe, is freely available for download at the following web address \url{http://web.pd.astro.it/mapelli/group.html#software}. An open access GitLab repository is available at \url{https://gitlab.com/mario.spera/SEVN}. 

In this paper we use \sevn{} as a stand-alone code for population synthesis calculations but it can also be coupled with a large variety of $N$-body codes. We have already included \sevn{} in the \starlab{} software environment \citep{portegieszwart2001,spera2016} and in an updated version of the \higpus{} code \citep{dolcetta2013}, called \higpusr{}. Preliminary access to the GitLab repository of \higpusr{} is available upon request through the email \url{mario.spera@live.it}.

\section{Results}
\label{sec:results}

\begin{figure*}
	\includegraphics[width=0.99\hsize]{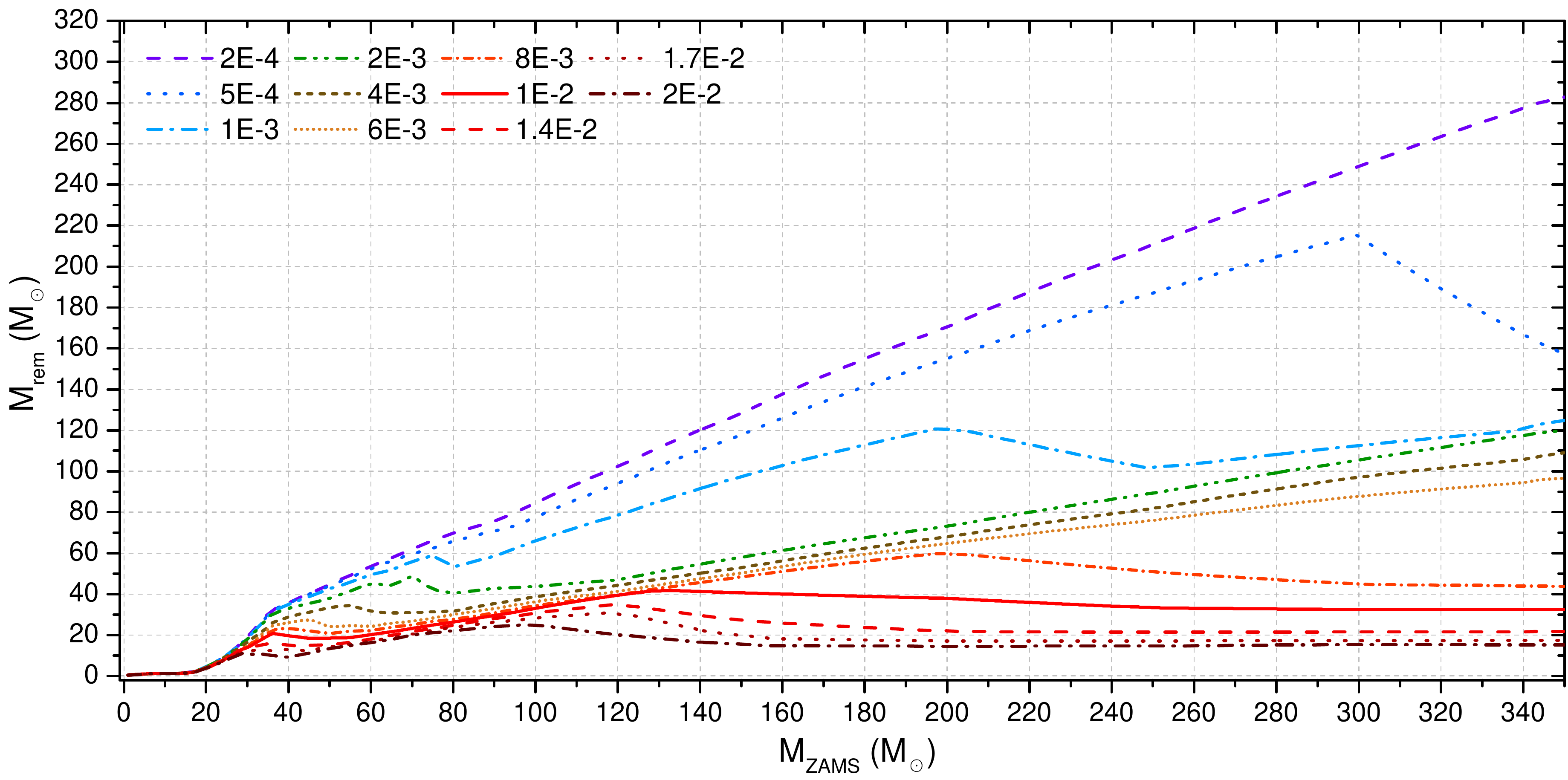}
	\caption{Mass of the compact remnant (\mrem{}) as a function of the ZAMS mass of the star (\mzams{}), derived with \sevn{}, without PPISNe and PISNe. From bottom to top: dash--dotted brown line: $Z=2.0\times 10^{-2}$; dotted dark orange line: $Z=1.7\times 10^{-2}$; dashed red line: $Z=1.4\times 10^{-2}$; solid red line: $Z=1.0\times 10^{-2}$; short dash-dotted orange line: $Z=8.0\times 10^{-3}$; short dotted light orange line: $Z=6.0\times 10^{-3}$; short dashed line: $Z=4.0\times 10^{-3}$; dash--double dotted line: $Z=2.0\times 10^{-3}$; dash--dotted light blue line: $Z=1.0\times 10^{-3}$; dotted blue line: $Z=5.0\times 10^{-4}$; dashed violet line: $Z=2.0\times 10^{-4}$.}
	\label{fig:mrem_mzams}
\end{figure*}

Fig. \ref{fig:mrem_mzams} shows the mass of the compact remnant as a function 
of the ZAMS mass of the progenitor star, for metallicity $Z$ ranging from $2.0\times 10^{-4}$ to $2.0\times 10^{-2}$. In \sevn{} and throughout this paper, we use $\zsun{}=0.01524$ for solar metallicity \citep{caffau2011}. Thus, the metallicity range in Fig. \ref{fig:mrem_mzams}  corresponds to $0.013-1.312\,{}\zsun$. Fig. \ref{fig:mrem_mzams} is an updated version of Fig.~6 of \citet{spera2015}, extending the maximum considered ZAMS mass from  $150\msun{}$ to $350\msun{}$ (using the evolutionary tracks from \citealt{chen2015}). 
Fig. \ref{fig:mrem_mzams} does not include the effect of PPISNe and PISNe. The maximum BH mass we obtain in this case is $\sim 280\msun{}$ at $Z = 2.0\times 10^{-4}$, from a progenitor star with $\mzams{}\simeq 350\msun{}$. 

In Fig. \ref{fig:mrem_mzams}, we adopt the delayed SN model. It is worth noting that, for stars with $\mzams{} \gtrsim 40\msun{}$, the mass of the BH does not depend on the adopted core-collapse SN explosion model but only on: (i) the effectiveness of stellar winds,  and (ii) the mass loss due to the escape of neutrinos at the onset of core collapse. Using the \parsec{} evolutionary models, all stars with $\mzams{}\gtrsim 40\msun{}$ undergo direct collapse, since they  have $\mcof{}\gtrsim 11\msun{}$.

\begin{figure*}
	\includegraphics[width=1.01\hsize]{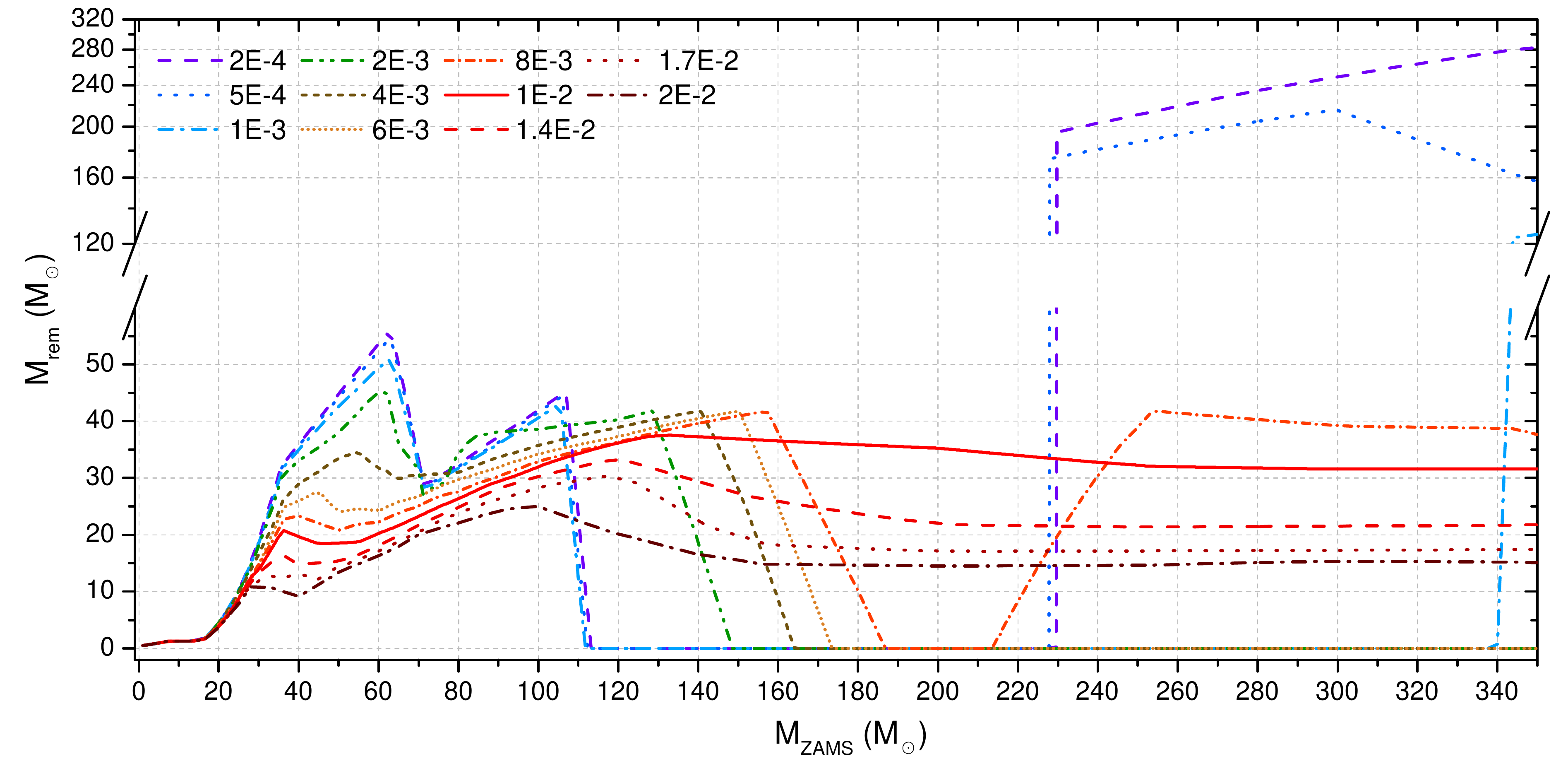}
	\caption{The same as Fig. \ref{fig:mrem_mzams} but with  PPISNe and PISNe. We have inserted a $y$-axis break between $65\msun{}$ and $120\msun{}$ because we have no  BHs in this mass range.}
	\label{fig:mrem_mzams_PISN}
\end{figure*}

Fig. \ref{fig:mrem_mzams_PISN} is the same as Fig. \ref{fig:mrem_mzams} but here we switched on PPISNe and PISNe. Both PPISNe and PISNe do not affect metal-rich stars. PPISNe first appear at $Z \leq{} 10^{-2}\simeq 0.7\zsun{}$ and their effects become significant at $Z \lesssim 2.0 \times 10^{-3}$, for stars with $60 \lesssim \mzams{}/\msun{} \lesssim 125$. From Fig.~\ref{fig:mrem_mzams_PISN} is apparent that the main effect of PPISNe is to lower the curves of Fig. \ref{fig:mrem_mzams} and to enhance the formation of BHs in the mass range  $30\lesssim \mrem{}/\msun{}\lesssim 50$. 

PISNe affect stars with metallicity $Z \leq{} 8.0 \times 10^{-3}\simeq 0.5\zsun{}$. Stars with $\mzams{} \gtrsim 210 \msun{}$, at $Z = 8.0\times 10^{-3}$, do not undergo the PISN because such VMSs are close to their Eddington's luminosity. Thus, they lose a larger fraction of mass than lighter stars and they cannot reach $\mhef{} \simeq 64\msun{}$ (i.e. the lower limit to activate the PISN mechanism).

Considering only stars with $\mzams{}\lesssim 200\msun{}$, the maximum BH mass we obtain with PPISNe and PISNe is $\sim 55\msun{}$ at $Z \simeq 2.0\times 10^{-4}\simeq 0.01\zsun{}$. Very massive BHs ($120-280$ \msun{}) can form from the direct collapse of VMSs with $\mzams{} \gtrsim 200\msun{}$ at $Z \lesssim 10^{-3}\simeq 0.07\zsun{}$. Such massive stars have $\mhef{}\gtrsim 135\msun{}$ and avoid PISNe. BHs with mass $\gtrsim{}120$ \msun{} are in the mass range of IMBHs.

 Another effect of PPISNe and PISNe is that there is a dearth of compact remnants in the mass range between $\sim 60\msun{}$ and $\sim 120\msun{}$. This is apparent in Fig. \ref{fig:mrem_mzams_PISN}, where we have inserted a $y$--axis break, corresponding to this range of BH masses. This result is in agreement with the mass gap found by \citet{belczynski2016} and W17. In Appendix \ref{app:C}, we compare our results with those obtained by \citet{woosley2017}.

In Appendix \ref{app:D}, we provide detailed tables with the values of the most relevant quantities that have been used to produce Figs~\ref{fig:mrem_mzams} and \ref{fig:mrem_mzams_PISN} (namely ZAMS mass, final mass, Helium core mass, CO mass, and remnant mass with and without PPISNe and PISNe) for three different metallicities ($Z=2.0\times 10^{-2}$, $Z=2.0\times 10^{-3}$, and $Z=2.0\times 10^{-4}$).

\begin{figure}
	\includegraphics[width=\columnwidth]{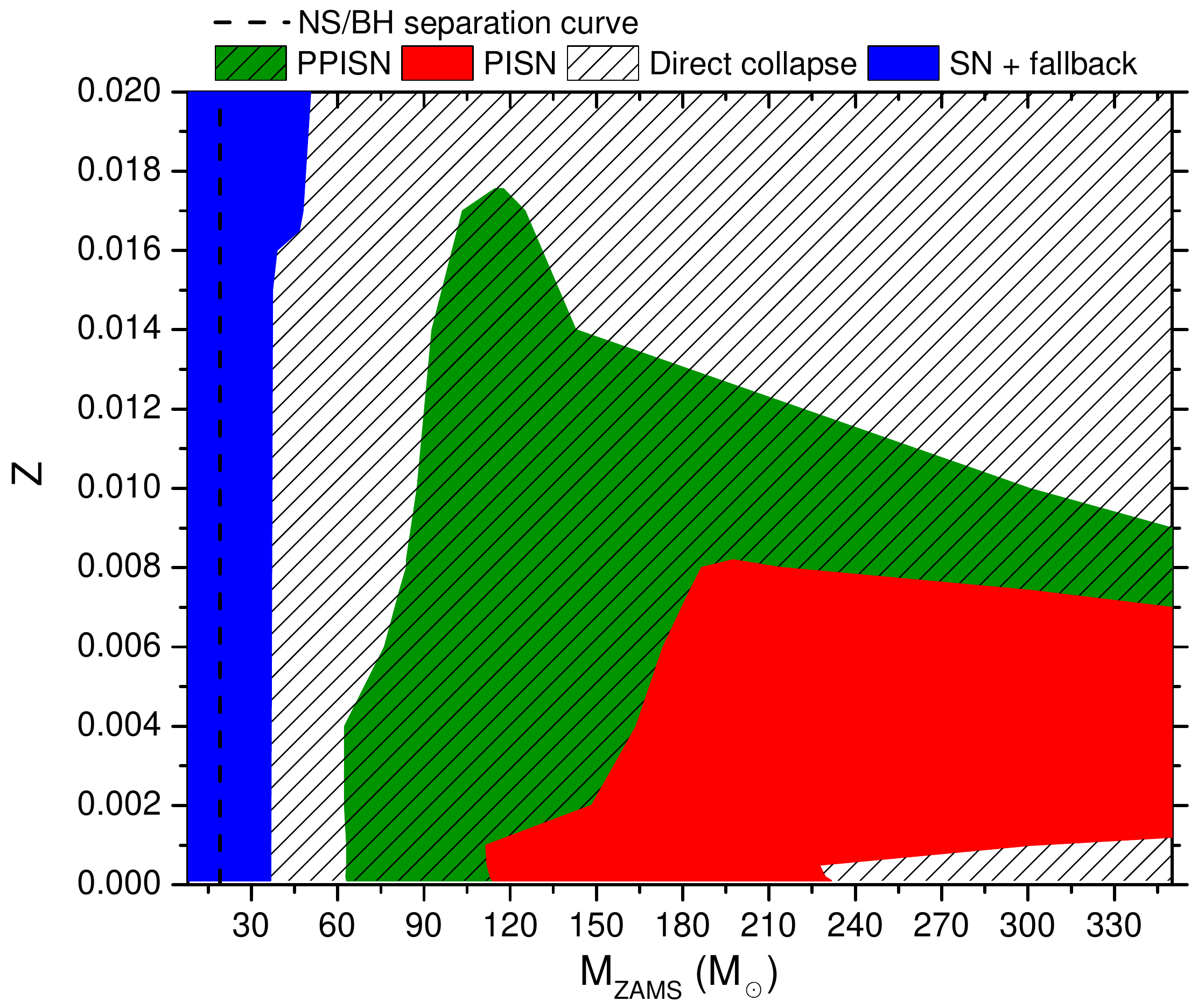}
	\caption{Regions of the $Z-\mzams{}$ plane where different SN mechanisms take place.  Stars in the blue area end their life through a core-collapse SN with fallback. PPISNe and PISNe occur in the green and in the red area, respectively. The hatched area indicates direct collapse. The vertical dashed line at $\mzams{} \simeq 19 \msun{}$ divides  stars that collapse into BHs ($\mzams{} \gtrsim 19 \msun{}$) from those forming NSs ($\mzams{} \lesssim 19 \msun{}$).}
	\label{fig:z_mzams}
\end{figure}

Fig. \ref{fig:z_mzams} shows the remnants of massive single stars in the $Z-\mzams{}$ plane. Fig. \ref{fig:z_mzams} is an updated version of Fig. 1 of \citet{heger2003} that shows the regions where PPISNe (green area) and PISNe (red area) occur, and the region where neutrino-driven SNe and fallback mechanism take place (blue area). The hatched area is where stars undergo direct collapse.
It is worth noting that all the stars that undergo the PPISN form compact remnants via direct collapse. The reason is that PPISNe enhance mass loss from the stars' external layers leaving the Carbon--Oxygen core unaffected. Thus, such stars have always $\mcof{}\gtrsim 11\msun{}$, which is the lower limit for a failed SN in the delayed mechanism. This Figure also shows that the lower limit in terms of \mzams{} for direct collapse depends on metallicity. This limit is $\sim 35\msun{}$ for $Z\leq 1.6\times 10^{-2}$ and it grows up to $\sim 50\msun{}$ at $Z=2.0\times 10^{-2}$.
{A difference with Fig. 1 of \citet{heger2003} is that in Fig. \ref{fig:z_mzams} we show the area where PPISNe occur. Furthermore, it is worth noting that we do not form NSs or BHs by fallback from stars with $\mzams{}\gtrsim 50\msun{}$. This is a consequence of the up-to-date prescriptions of stellar winds implemented in the PARSEC code that predict $\mcof{}\gtrsim 11\msun{}$ for $\mzams{}\gtrsim 50\msun{}$, at any metallicity.}

\section{Discussion}
In this section we discuss the impact of PISNe and PPISNe on the detected GW events and on the formation of IMBHs.

\subsection{Formation environment of the gravitational-wave detections}

To test the impact of PISNe and PPISNe on GW events, we performed two sets of population synthesis simulations which we refer to as \simA{} and \simB{}. Each set consists of 30 simulations at different metallicity in the range between $Z=1.0\times 10^{-4}$ and $Z=2.0\times 10^{-2}$. The number of stars in each run is $10^8$. We verified that this number of objects is high enough to filter out statistical fluctuations in the analysis. All stars evolve in isolation for $\Delta t=150$ Myr, which is a sufficiently long time to ensure that all the BHs have formed. The initial masses of the stars are sampled from a broken power--low mass function \citep{kroupa2001},  with a range $0.1<\mzams{}/\msun{}<150$, and with slopes $\alpha_{\mathrm{1}} = 1.3$ for $0.1\leq m/\msun{}<0.5$ and $\alpha_{\mathrm{2}} = 2.3$ for $0.5\leq m/\msun{}\leq 150$. The only difference between \simA{} and \simB{} is that PPISNe and PISNe have been switched off in \simB{}. 

\begin{figure}
	\includegraphics[width=\columnwidth]{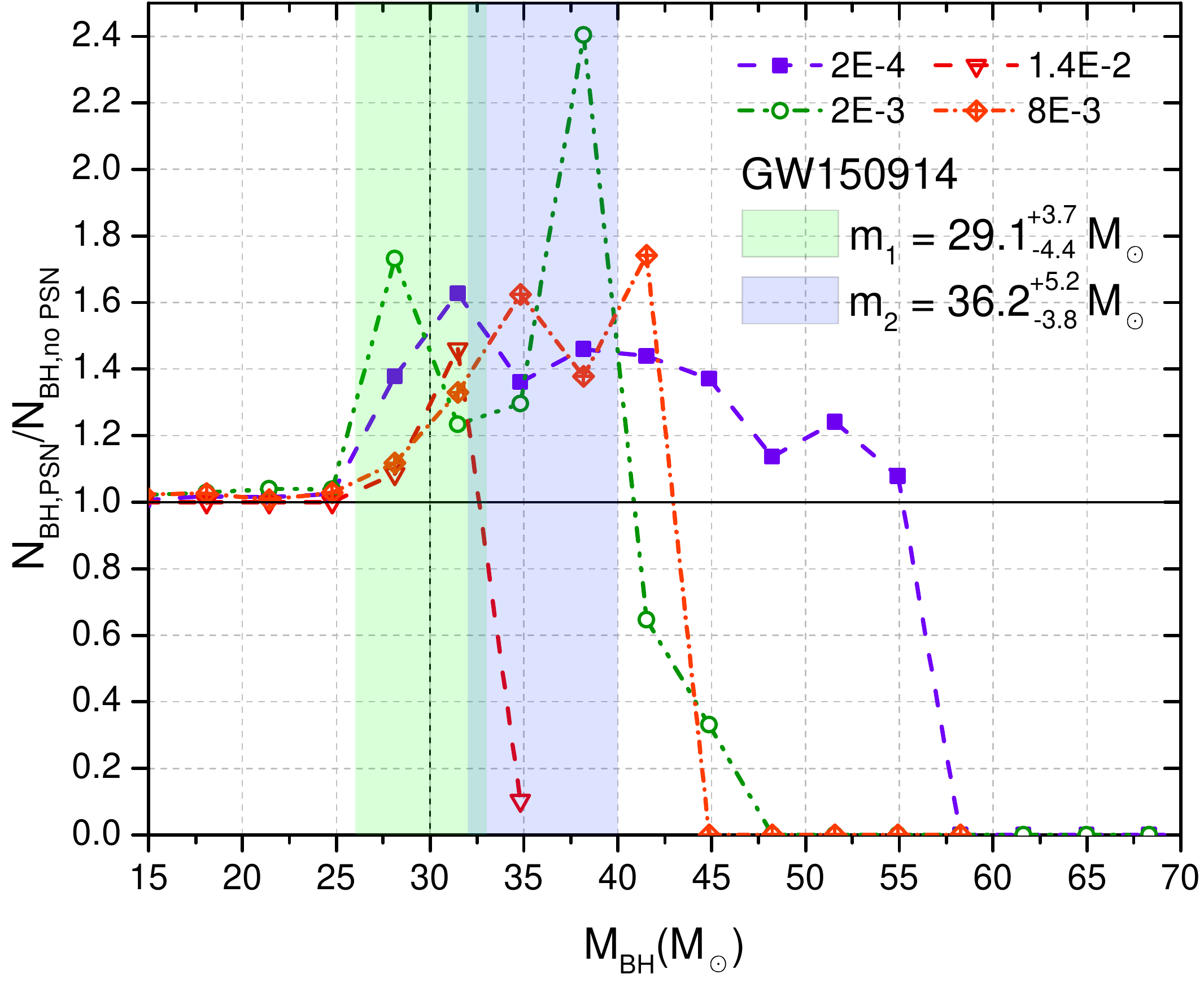}
	\caption{Ratio between the number of BHs obtained in the population synthesis simulations of \simA{} and \simB{}, respectively, as a function of the BH mass, and for different metallicity. To construct this Figure, all the BHs in the mass range $3\div 80\msun{}$ have been  gathered into 20 mass bins. The abscissa of each point  is the middle point of each mass bin.  The green and blue shaded areas represent the masses of the two BHs of GW150914 with the associated uncertainties. Line types are the same as in Fig. \ref{fig:mrem_mzams} and indicate different metallicities.}
	\label{fig:fraction_mbh}
\end{figure}

Fig. \ref{fig:fraction_mbh} shows the ratio between the number of BHs we obtained from the simulations of \simA{} and that obtained from the simulations of \simB{}, as a function of the BH mass, for different values of metallicity. It is apparent that PPISNe favour the formation of BHs with masses between $\simeq 25\msun{}$ and $\simeq 50\msun{}$. This result applies to a wide range of metallicities ($2.0\times 10^{-4}\lesssim Z \lesssim 1.4\times 10^{-2}$). For $Z\gtrsim2.0\times 10^{-2}$ neither PPISNe nor PISNe are effective so \simA{} and \simB{} generate the same results. In Fig. \ref{fig:fraction_mbh}, we also show the masses of the two BHs of GW150914 and their associated uncertainties. In \simA{} we form from $\sim 1.5$ to $\sim 2.4$ times more GW150914--like BHs than in the simulations of \simB{}. This result also applies to the primary BH of GW170104. In contrast, the effect of PISNe is to reduce the number of massive BHs in the runs of \simA{}. The maximum BH mass depends on metallicity and it is $\simeq 35$, 45, 48, 60 \msun{} for $Z\simeq 1.4\times 10^{-2}$, $8.0 \times 10^{-3}$, $2.0\times 10^{-3}$, and  $2.0\times 10^{-4}$, respectively.

From our runs, we can estimate the most probable metallicity of the formation environment of GW detections. For GW150914, in both sets of runs and for each value of metallicity, we counted the number of BHs ($n_1$) with mass $26\leq m_{\mathrm{BH}}/\msun{}\leq 33$ and those ($n_2$) with mass $32\leq m_{\mathrm{BH}}/\msun{}\leq 40$. These two mass ranges are the confidence intervals for the BH masses of GW150914 (see e.g. \citealt{abbott2016d}). We divided the numbers $n_1$ and $n_2$ by the total number of simulated stars ($10^8$) to obtain the relative probability ($P(n_1)$ and $P(n_2)$, respectively) to form the two BHs in our simulations. The final probability to obtain a pair of GW150914--like BHs is then $P(n_1,n_2)\equiv \mathrm{min}\left(P(n_1), P(n_2)\right)$. Similarly, we construct the same quantity for GW170104, GW151226 and LVT151012. We stress that our definition of $P(n_1,n_2)$ contains severe approximations, because it does not account for either binary evolution processes or dynamical interactions.

\begin{figure}
	\includegraphics[width=\columnwidth]{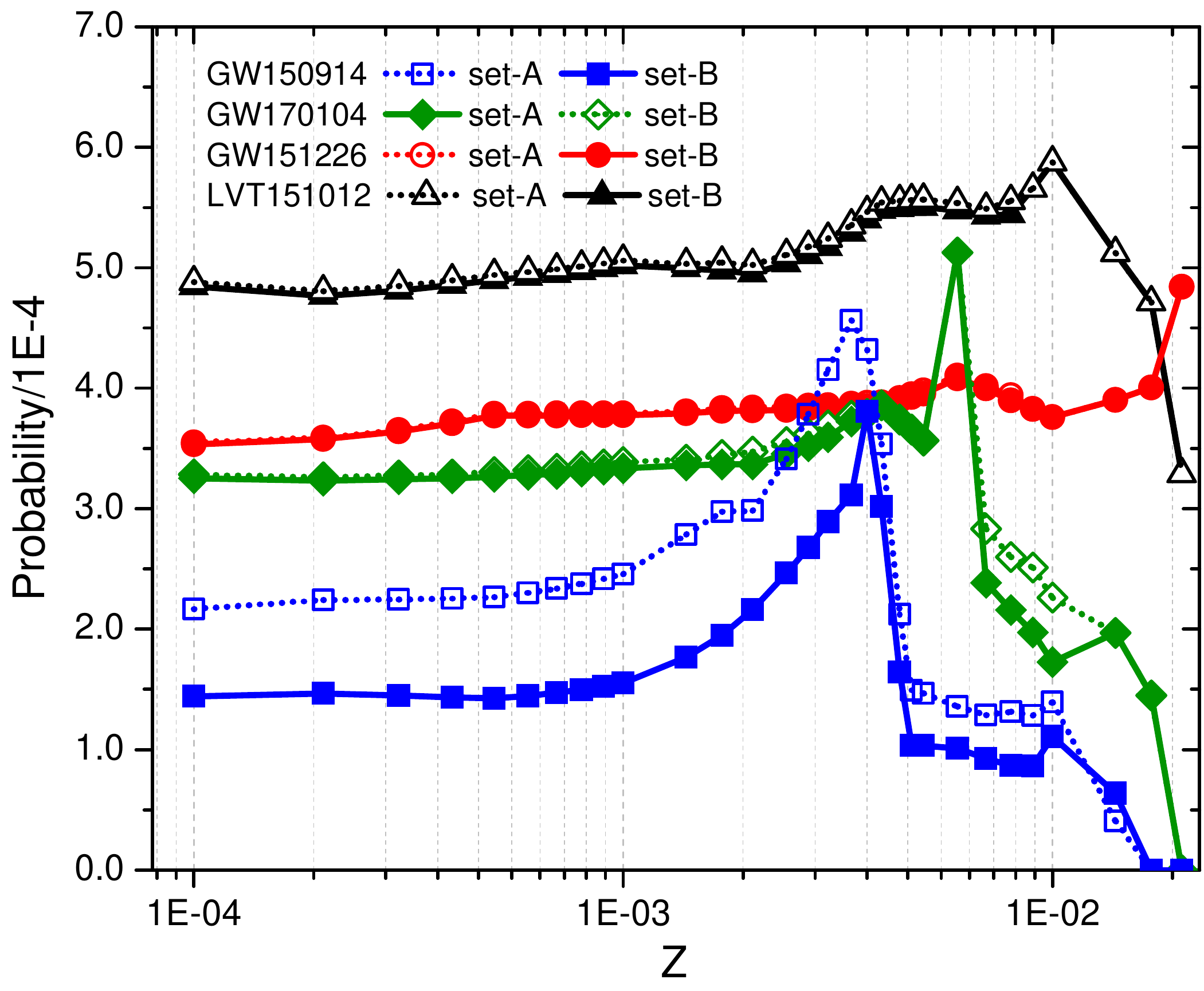}
	\caption{Relative probability, normalized to $10^{-4}$, to obtain BH pairs like those observed by aLIGO in its observing runs, as a function of metallicity. Empty symbols connected through dashed lines show the results obtained from the simulations of \simA{}, while filled symbols connected with solid lines refer to simulations of \simB{}. We use blue squares for GW150914, green rhombi for GW170104, red circles for GW151226, and black triangles for LVT151012.}
	\label{fig:probability_z}
\end{figure}

Fig. \ref{fig:probability_z} shows  $P(n_1,n_2)$, normalized to $10^{-4}$, as a function of metallicity, for the four GW events, obtained from the simulations of \simA{} and \simB{}. Fig. \ref{fig:probability_z} indicates that in both sets of simulations, the probability curve of GW150914 favours low--metallicity with respect to high--metallicity environments. The curve peaks at about $3\times 10^{-3}\lesssim Z \lesssim 4\times 10^{-3}$ and rapidly decreases for $Z>4\times 10^{-3}$, becoming zero for $Z\gtrsim 1.7\times 10^{-2}$. At $Z\gtrsim 1.7\times 10^{-2}$, stellar winds become very effective, preventing the formation of BHs with mass above $\sim 25\msun{}$ (see Fig. \ref{fig:mrem_mzams_PISN}). From Fig. \ref{fig:probability_z} we argue that the progenitors of GW150914 likely formed in a metal--poor environment with metallicity $Z \lesssim 4.0\times 10^{-3}\simeq 0.3\,{}\zsun{}$. This result is in agreement with the findings of \citet{abbott2016b} and \citet{belczynski2016nat}. 

The probability curve of GW170104 is similar to that of GW150914, We obtain a maximum formation probability at about $6.0\times 10^{-3}Z\lesssim Z \lesssim 7.0\times 10^{-3}$. The curve rapidly decreases at higher metallicity, becoming zero at $Z = 2.0\times 10^{-2}$, and flattens for $Z\lesssim 5.0 \times 10^{-3}$. Thus, we can argue that GW170104 likely formed in a metal--poor environment with metallicity $Z\lesssim 7.0\times 10^{-3} \simeq 0.5\zsun{}$.

The BHs observed in GW151226  are lighter ($m_1=14^{+8}_{-4}\msun{}$, $m_2=7.5^{+2}_{-2}\msun{}$). Since our models predict the formation of such BHs at every metallicity (see Fig. \ref{fig:mrem_mzams_PISN}), the probability curve of GW151226 shown in Fig. \ref{fig:probability_z} (red circles) is almost flat. High values of metallicity ($Z\gtrsim 10^{-2}$) are slightly favoured because, in metal--rich environments, all progenitor stars form BHs with masses below $\sim 25\msun{}$. Furthermore, the probability curve obtained from \simA{} and \simB{} are overlapped. This happens because PPISNe and PISNe do not affect the formation of BHs with mass below $\sim 25\msun{}$ (see Fig. \ref{fig:mrem_mzams_PISN}). Fig. \ref{fig:probability_z} also shows the probability curve of LVT151012 ($m_1=23^{+18}_{-6}\msun{}$, $m_2=13^{+5}_{-5}\msun{}$).

\begin{figure*}
	\includegraphics[width=\hsize]{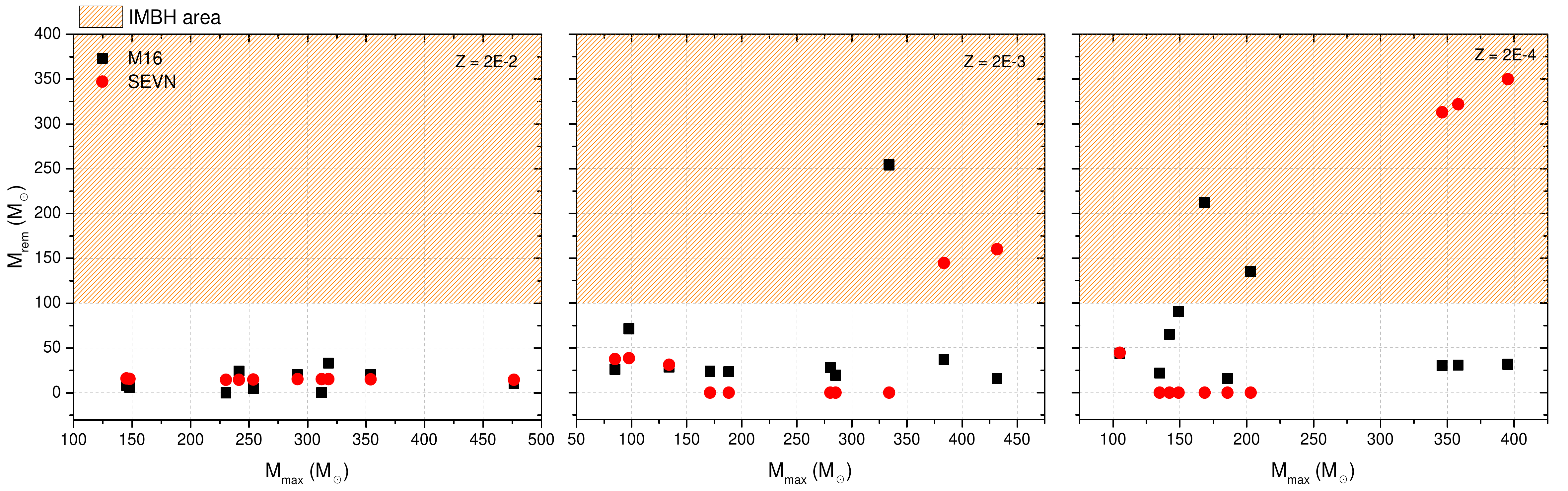}
	\caption{Mass of the compact remnant as a function of the maximum mass of the principal collision product (PCP). Black squares: simulations of \citet{mapelli2016}; red circles: this work. Left-hand panel: $Z=2.0\times 10^{-2}$; middle panel: $Z=2.0\times 10^{-3}$; right-hand panel: $Z=2.0\times 10^{-4}$. The orange shaded region is the mass range of IMBHs ($>100\msun{}$). }
	\label{fig:rem_max_PCP_3panels}
\end{figure*}

\subsection{Formation of intermediate--mass black holes}
The existence of IMBHs (mass between $\sim 100\msun{}$ and $\sim 10^{5}\msun{}$) is still matter of debate \citep{farrell2009,strader2012,lutzgendorf2013,lanzoni2013,baumgardt2017,kiziltan2017,zocchi2017}. 

From a theoretical point of view, several possible formation mechanisms have been proposed (e.g. \citealt{giersz2015} and references therein). One of them is the so called runaway collision scenario: in a dense stellar system, a massive star  (with mass $> 100\msun{}$) may form through a series of collisions and then may directly collapse into an IMBH  \citep{portegieszwart2002,portegieszwart2004}. Recently, \cite{mapelli2016} has studied the impact of stellar winds on the formation of IMBHs from runaway collisions. \cite{mapelli2016} performed a set of direct $N$-body simulations of star clusters by means of the \starlab{} software environment \citep{portegieszwart2001}. Stellar evolution was modified to include metallicity dependence and recent prescriptions for stellar winds, as described in \cite{mapelli2013}.  \cite{mapelli2016} considered three different metallicities ($Z=2.0\times 10^{-2}$, $Z=2.0\times 10^{-3}$, and $Z=2.0\times 10^{-4}$), performing, for each metallicity, ten different realizations of the same cluster (with $10^5$ stars). The evolution of the principal collision product (PCP), defined as the product of the first collision that occurs in a simulated star cluster, is tracked in each simulation. 

\cite{mapelli2016} finds that no IMBH can form at solar metallicity, because of the enhanced mass loss, whereas runaway collisions might still produce IMBHs at metallicity $\lesssim{}0.1$ \zsun. The simulations of \cite{mapelli2016} do not include PPISNe and PISNe and adopt mass loss prescriptions for VMSs that are extrapolated from formulas derived for ``ordinary'' massive stars ($\sim{}30-150$ M$_\odot$, \citealt{mapelli2013}). Our aim is to check what is the impact of (i) the new stellar tracks for VMSs and of (ii) PPISNe and PISNe on the runaway collision products simulated by \cite{mapelli2016}. Thus, we have re-run each PCP simulated by \cite{mapelli2016}. We use \sevn{} including PPISNe and PISNe as described in Section~\ref{sec:method}. We start the evolution of each PCP with \sevn{} from the time when it reaches its maximum mass (see Figs. 2, 3 and 4 of \citealt{mapelli2016}). In the new evolution of the PCP, we apply self-consistent stellar evolutionary models \citep{chen2015} for stars with mass  $\leq{}350\msun{}$ and  we use a linear extrapolation of the curves shown in Fig. \ref{fig:mrem_mzams_PISN} for more massive stars. The main approximation of our approach is that we do not account for collisions occurring after the PCP has reached its maximum mass.

Fig. \ref{fig:rem_max_PCP_3panels} compares the compact-object masses  obtained with \sevn{}  and those reported in \citet{mapelli2016}, as a function of the maximum mass reached by the PCP. The three panels are for three different values of metallicity. 
For $Z=2.0\times 10^{-2}$, the two sets of points distribute approximately in the same $\mrem{}$ range. The mass of the compact object formed by the PCP is always below $\sim 35\msun{}$. At this high metallicity, mass loss through stellar winds is  effective so even very massive stars  may end their lives with no more than $\sim 40\msun{}$, preventing the formation of IMBHs. At lower metallicity, we find several differences between \sevn{} and \cite{mapelli2016}. 

At $Z=2.0\times 10^{-3}$, one PCP in \cite{mapelli2016} reaches a maximum mass of $\sim 335\msun{}$. In \cite{mapelli2016}, this object undergoes direct collapse, forms a massive BH with mass $\sim 210\msun{}$, and becomes an object of mass $\sim 250\msun{}$, after a further collision with a main--sequence star. In contrast, when using \sevn{}, this PCP undergoes a PISN and does not leave a compact object. At the same metallicity,  the \sevn{} prescriptions allow other two PCPs (with $M_{\mathrm{max}}\simeq 375\msun{}$ and $M_{\mathrm{max}}\simeq 430\msun{}$, respectively) to form two IMBHs with mass $\mrem{}\simeq 150\msun{}$, while the extrapolated fitting formulas used in \cite{mapelli2016} predict the formation of two lighter BHs (mass below $\sim 40\msun{}$). 

 At $Z=2.0\times 10^{-4}$ (right-hand panel) two PCPs of \cite{mapelli2016} form IMBHs with $M_{\mathrm{rem}}\simeq 215\msun{}$ and $M_{\mathrm{rem}}\simeq 140\msun{}$, respectively. Using \sevn{}, these two PCPs do not collapse into IMBHs since they undergo a PISN. Similarly, our new models predict that the three PCPs with $M_{\mathrm{max}}\gtrsim 340\msun{}$ avoid a PISN and collapse into  IMBHs, with masses above $\sim 300\msun{}$.  

Fig. \ref{fig:rem_max_PCP_3panels} shows that PPISNe, PISNe and up-to-date stellar evolution models for VMSs \citep{chen2015} can significantly affect the evolution of the PCP. This suggest that a set of self--consistent $N-$body simulations including \sevn{} is absolutely necessary to get more insights on the runaway collision mechanism, and will be performed in a follow-up paper. 

The bottom line of our preliminary study is that no IMBHs can form at solar metallicity from runaway collisions (in good agreement with \citealt{mapelli2016}), while $\sim{}20-30$ per cent of runaway collision products can collapse to IMBHs at $Z\leq{}0.002$ (\citealt{mapelli2016} predicts that $\sim{}10-20$ per cent of runaway collision products can form IMBHs at $Z\leq{}0.002$).

\section{Summary}

We described an updated version of our population synthesis code, \sevn{}, where we included an analytical prescription for PPISNe and PISNe (derived from \citet{woosley2017}), and up-to-date stellar evolution tracks for VMSs, with \mzams{} up to $350\msun{}$ \citep{chen2015}. The new version of \sevn{} is publicly available and can be downloaded from \url{http://web.pd.astro.it/mapelli/group.html#software} or \url{https://gitlab.com/mario.spera/SEVN}.

We used the new version of \sevn{} to study the BH mass spectrum at different metallicities, ranging from $Z=2.0\times 10^{-4}$ to $Z=2.0\times 10^{-2}$. We find that  the effect of PPISNe becomes significant for $Z\leq 2.0\times 10^{-3}\simeq 0.1\zsun{}$, for stars with $60\le{}\mzams{}/\msun{}\le{}125$. PISNe are  effective in the range $1.0\times 10^{-3}\leq Z \leq 8.0 \times 10^{-3}$ for stars with $150\leq \mzams{}/\msun{} \leq 350$ (the lower mass limit slightly depending on metallicity). For $Z\lesssim 1.0\times 10^{-3}\simeq 0.07\zsun{}$, VMSs do not undergo a PISN and collapse directly to IMBHs with mass $\gtrsim 200\msun{}$.

Moreover, PPISNe and PISNe enhance the formation of BHs in the mass range $30\leq\mrem{}/\msun{}\leq 50$, while preventing the formation of compact remnants with mass $60\leq \mrem{}/\msun{}\leq 120$.  This implies that PPISNe and PISNe favour the formation of BHs with mass between $25\msun{}$ and $50\msun{}$, i.e. in the mass range of the GW150914 ($m_1=36.2^{+5.2}_{-3.8}\msun{}$ and $m_2=29.1^{+3.7}_{-4.4}\msun{}$). From our simulations, we estimated that GW150914 and GW170104 likely formed in a metal-poor environment with metallicity $Z\leq 0.3\zsun{}$ (see also \citealt{abbott2016b} and \citealt{belczynski2016nat}), and $Z\leq 0.5\zsun{}$, respectively.

Finally, we discuss the formation of IMBHs from VMSs. We studied the impact of PPISNe and PISNe on the runaway collision products formed in the direct $N$-body simulations of \citet{mapelli2016}, who do not include the effect of PPISNe and PISNe. In our simulations, no IMBHs form from runaway collisions of metal-rich stars ($Z=0.02$), in agreement with \citet{mapelli2016}. In metal-poor star clusters, we find that $\sim{}20-30$ per cent of runaway collision products collapse to IMBHs (\citet{mapelli2016} finds $\sim{}10-20$ per cent). 
There are significant differences in the fate of each single collision product between this paper and \citet{mapelli2016}. These differences arise from the different recipes adopted for the evolution of VMSs, from the effect of PPISNe and PISNe, but also from the fact that we do not integrate the dynamical evolution of the collision products. Thus, in a forthcoming paper we will perform a set of self-consistent $N$-body simulations including the new version of \sevn{} to shed light on the formation of IMBHs from VMSs in star clusters.

\section{Acknowledgments}	
We thank the anonymous referee for their comments, which helped us to improve the manuscript. We thank Alessandro Bressan for his invaluable suggestions and useful discussions. MM and MS acknowledge financial support from the Italian Ministry of Education, University and Research (MIUR) through grant FIRB 2012 RBFR12PM1F and from INAF through grant PRIN-2014-14. MM acknowledges support from the MERAC Foundation. Part of the Numerical calculations have been made possible through a CINECA-INFN agreement, providing access to resources on GALILEO and MARCONI at CINECA.

	\bibliographystyle{mnras}
	\bibliography{SperaM_MapelliM_bibtex}

\appendix

\section{A new method to interpolate stellar evolution tracks in \sevn{}}
\label{app:B}

\begin{figure}
	\includegraphics[width=\hsize]{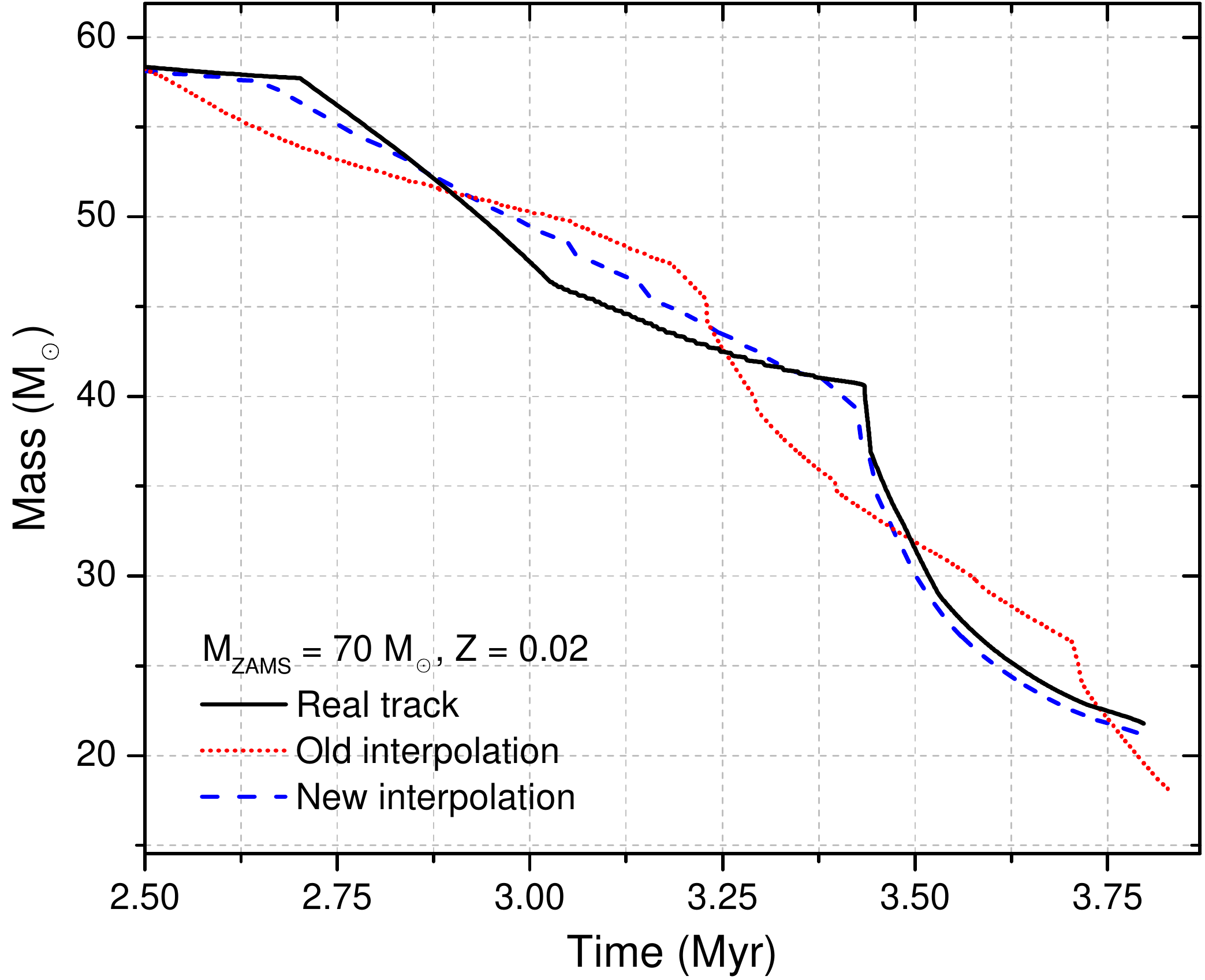}
	\caption{Time evolution of the mass of a star with $\mzams{}=70\msun{}$ at $Z=2.0\times 10^{-2}$. The black solid curve is obtained using the \parsec{} stellar evolution code. Dotted red line: interpolation using the old version of \sevn{}; dashed blue line: new version of \sevn{}. Both the interpolated curves are obtained using the pre-evolved tracks of two stars with $M_1=60\msun{}$ and $M_2=80\msun{}$, respectively.}
	\label{fig:mass_interp}
\end{figure}

\begin{figure}
	\includegraphics[width=\hsize]{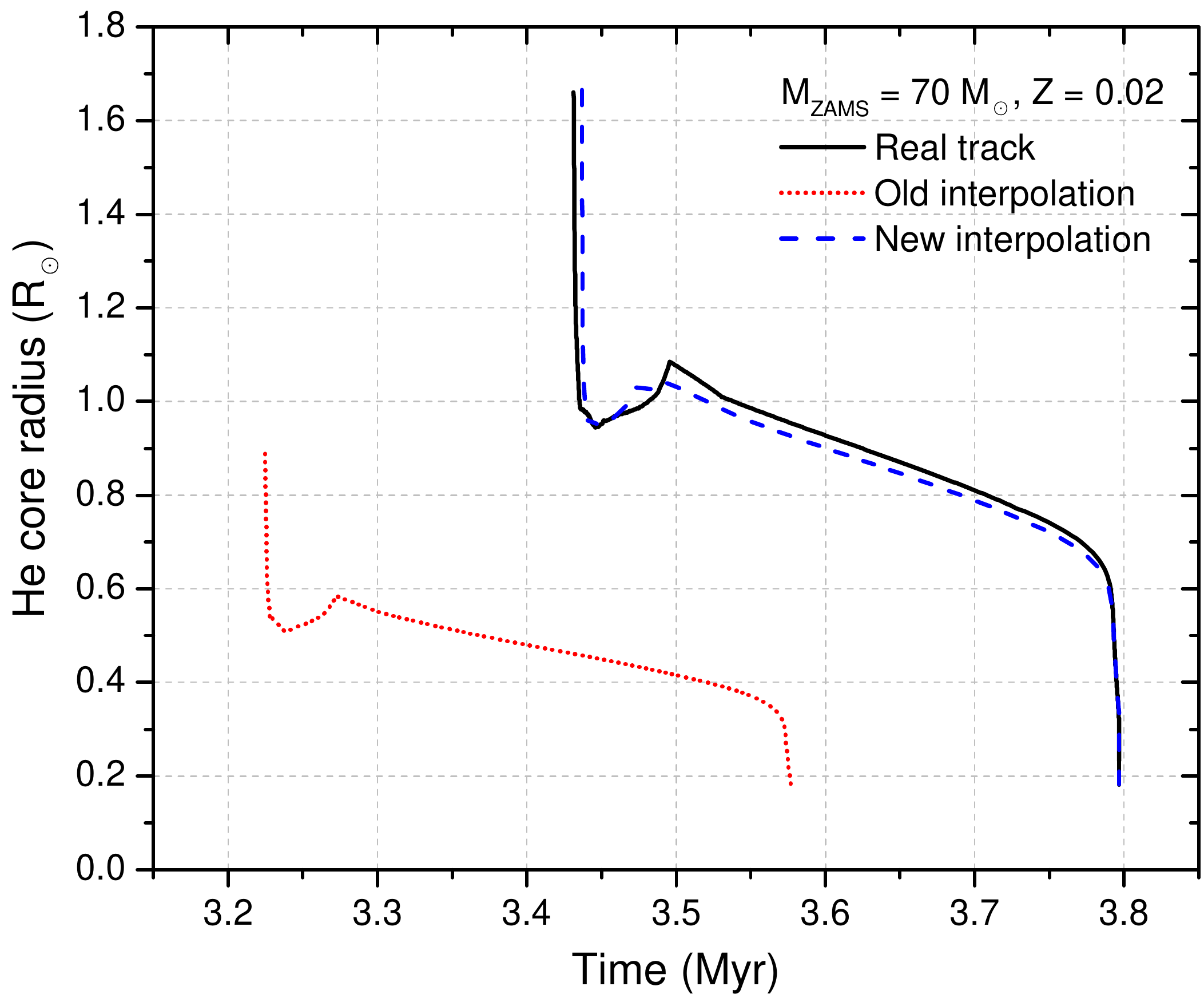}
	\caption{Same as Fig. \ref{fig:mass_interp} but for the time evolution of the Helium core radius.}
	\label{fig:Hecore_interp}
\end{figure}

The old version of \sevn{} evolves a star with $M_{\mathrm{ZAMS,s}}=M_s$ and metallicity $Z_s$ by interpolating the pre-evolved tracks of two stars with $M_{\mathrm{ZAMS,1}}=M_1=M_s-\Delta M$ and $M_{\mathrm{ZAMS,2}}=M_2=M_s+\Delta M$, respectively, where $\Delta M$ is the step of the mass grid in the input tables. The mass of the star at time $t = \widetilde{t}$ is obtained by linear interpolation:
\begin{equation}
M_s\left(t=\widetilde{t}\right) = \alpha_1 M_1\left(t=\widetilde{t}\right) + \alpha_2 M_2\left(t=\widetilde{t}\right)
\end{equation}
where $\alpha_1 \equiv \frac{M_2-M_s}{M_2-M_1}$ and $\alpha_2 \equiv \frac{M_s-M_1}{M_2-M_1}$.

Instead, the new version of \sevn{} uses the following algorithm. To evolve the star $s$ at time $t = \widetilde{t}$, the new version of \sevn{} evaluates the value $p=\dfrac{\widetilde{t}}{t_{\mathrm{life,s}}}$, where $t_{\mathrm{life,s}}$ is the lifetime of the star $s$. $M_s\left(t=\widetilde{t}\right)$ is calculated as
\begin{equation}
M_s\left(t=\widetilde{t}\right) = \beta_1 M_1\left(t=\widetilde{t_1}\right) + \beta_2 M_2\left(t=\widetilde{t_2}\right)
\end{equation}
where $\widetilde{t_1}\equiv pt_{\mathrm{life,1}}$, $\widetilde{t_2}\equiv pt_{\mathrm{life,2}}$, $\beta_1\equiv\frac{M_1\left(M_2-M_s\right)}{M_s\left(M_2-M_1\right)}$, and $\beta_2\equiv\frac{M_2\left(M_s-M_1\right)}{M_s\left(M_2-M_1\right)}$. This new approach ensures that the interpolating stars are in the same stellar phase of the target star $s$, and that the values of their physical parameters are appropriately weighted. This allows us to include less points in the input tables while reducing the interpolation error, as shown by Figs.  \ref{fig:mass_interp} and \ref{fig:Hecore_interp}.

These figures show a comparison between the new and the old interpolation method to approximate the temporal evolution of the mass (Fig.  \ref{fig:mass_interp}) and the Helium core radius (Fig. \ref{fig:Hecore_interp}) of a star with mass $\mzams{}=70\msun{}$ and metallicity $Z=2.0\times 10^{-2}$. It is apparent that the new interpolation method works much better than the previous one, especially for estimating the values of the physical parameters that are important for evolved stars (such as the Helium core radius).

It is worth noting that the new version of \sevn{} can also interpolate tables at different metallicity. That is, if the pre-evolved tracks are not available for the metallicity $Z_s$, \sevn{} first evaluates $M_s\left(t=\widetilde{t},Z=Z_1\right)$ and $M_s\left(t=\widetilde{t},Z=Z_2\right)$, where $Z_1\equiv Z_s - \Delta Z$, $Z_2\equiv Z_s + \Delta Z$, and $\Delta Z$ is step in the metallicity grid. Then, the value $M_s\left(t=\widetilde{t}, Z_s\right)$ is
\begin{equation}
M_s\left(t=\widetilde{t}, Z_s\right) = \gamma_1 M_s\left(t=\widetilde{t_1}, Z=Z_1\right) + \gamma_2 M_s\left(t=\widetilde{t_2}, Z=Z_2\right)
\end{equation}
where $\gamma_1 \equiv \frac{Z_s-Z_1}{Z_2-Z_1}$ and $\gamma_2 \equiv \frac{Z_2-Z_s}{Z_2-Z_1}$.

\section{The fitting formula for PPISNe and PISNe}
\label{app:A}
The new version of \sevn{} includes a fitting formula to derive \mrem{} accounting for PPISNe and PISNe. This formula has been obtained by fitting the masses of the compact remnants shown in Tab. 1 and Tab. 2 of \citet{woosley2017} as a function of the final Helium mass fraction and of the Helium core mass \mhef{} of the stars. Using the following fitting formulas we get the parameter \ap{} so that $\mrem{} = \ap{}M_{\mathrm{rem, no\,psn}}$ where $M_{\mathrm{rem, no\,psn}}$ is the mass of the compact remnant we would obtain without PPISN/PISN (see equation~\ref{eq:ap}). First, we define the following quantities

\begin{equation}
\begin{split}
\mathcal{F} &\equiv \dfrac{\mhef{}}{\mfin{}}\\
\mathcal{K} &\equiv 0.67000\mathcal{F}+0.10000\\
\mathcal{S} &\equiv 0.52260\mathcal{F}-0.52974.
\end{split}
\end{equation}
We then express \ap{} as a function of $\mathcal{F}$, $\mathcal{S}$, $\mathcal{K}$ and \mhef{}:

\begin{equation}
\ap{} =
\begin{cases}
1 \\
\text{\hspace{15 pt} if } \mhef{}\leq 32\msun{}, \,\, \forall \mathcal{F}, \,\, \forall \mathcal{S}\\

0.2\left(\mathcal{K}-1\right)\mhef{}+0.2\left(37-32\mathcal{K}\right)\\
\text{\hspace{15 pt} if } 32<\mhef{}/\msun{}\leq 37, \,\, \mathcal{F}<0.9, \,\, \forall \mathcal{S}\\

\mathcal{K}\\
\text{\hspace{15 pt} if } 37<\mhef{}/\msun{}\leq 60, \,\, \mathcal{F}<0.9, \,\, \forall \mathcal{S}\\

\mathcal{S}\left(\mhef{}-32\right) + 1\\
\text{\hspace{15 pt} if } \mhef{}\leq 37\msun{}, \,\, \mathcal{F}\geq 0.9, \,\, \forall \mathcal{S}\\

5\mathcal{S}+1\\
\text{\hspace{15 pt} if } 37<\mhef{}/\msun{}\leq 56, \,\, \mathcal{F}\geq 0.9,\,\, \mathcal{S}<0.82916\\

\left(-0.1381\mathcal{F}+0.1309\right)\left(\mhef{}-56\right)+0.82916\\
\text{\hspace{15 pt} if } 37<\mhef{}/\msun{}\leq 56, \,\, \mathcal{F}\geq 0.9,\,\, \mathcal{S}\geq 0.82916\\

-0.103645\mhef{}+6.63328\\
\text{\hspace{15 pt} if } 56<\mhef{}/\msun{}< 64, \,\, \mathcal{F}\geq 0.9, \,\, \forall \mathcal{S}\\

0\\
\text{\hspace{15 pt} if } 64\leq\mhef{}/\msun{}< 135, \,\, \forall \mathcal{F},\,\, \forall \mathcal{S}\\

1\\
\text{\hspace{15 pt} if } \mhef{}\geq 135\msun{}, \,\, \forall \mathcal{F}, \,\, \forall \mathcal{S}.\\
\end{cases}
\end{equation}

\section{Comparison with W17}
\label{app:C}
In this appendix we compare our results with those presented in \citet{woosley2017}.

\begin{figure}
	\includegraphics[width=\hsize]{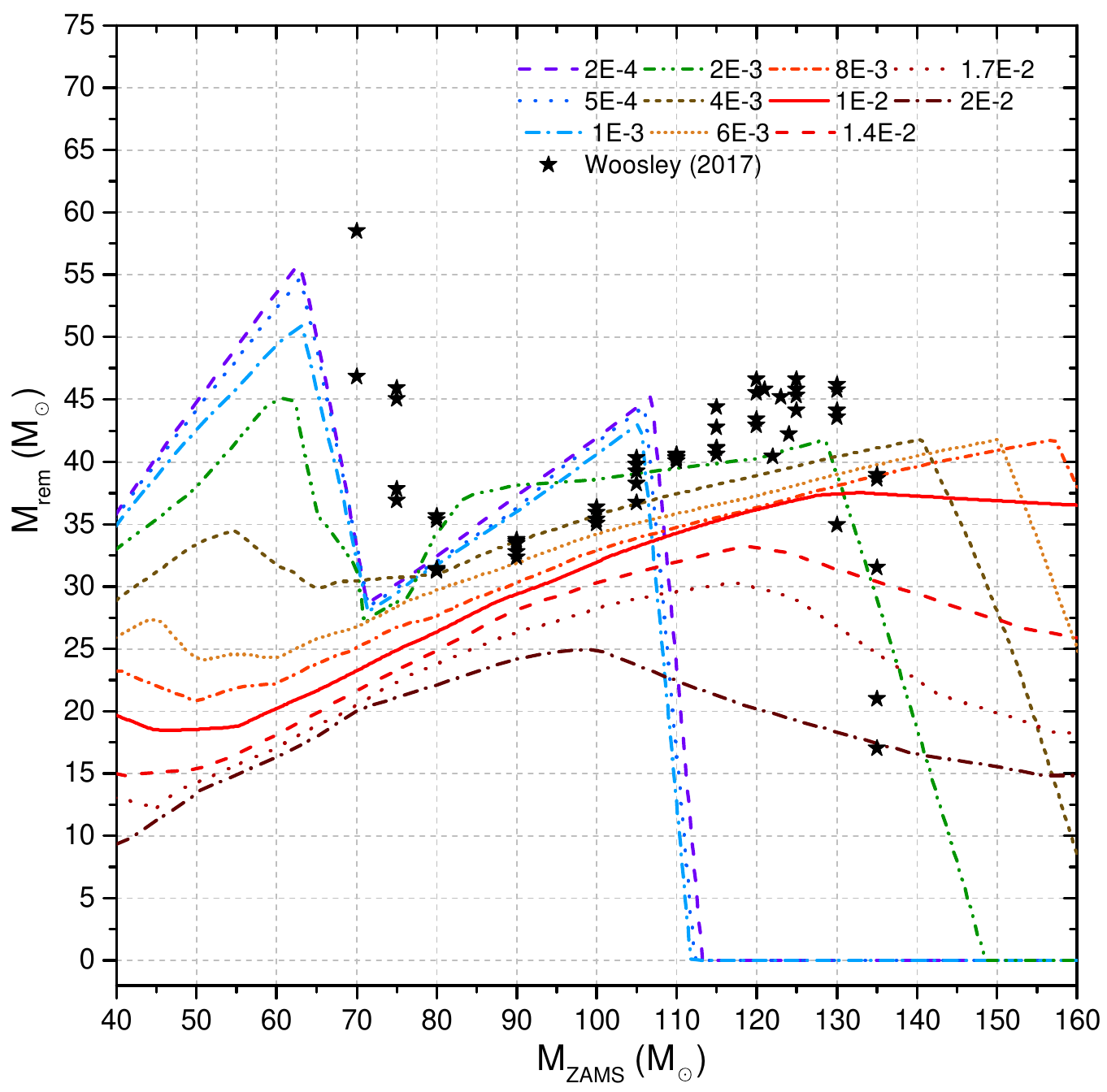}
	\caption{A detail of Fig. \ref{fig:mrem_mzams_PISN} that shows the ranges $40\leq \mzams{}/\msun{} \leq 160$ and $0\leq \mrem{}/\msun{} \leq 75$. Black points are the compact remnant masses taken from Table 2 of \citet{woosley2017}.}
		\label{fig:mrem_mzams_woosley}
	\end{figure}
	
	Fig. \ref{fig:mrem_mzams_woosley} is a detail of Fig. \ref{fig:mrem_mzams_PISN} where we also plot the compact remnant masses taken from Table 2 of W17 (black points). W17 explores the evolution of stars at $Z\simeq 0.1\zsun{}$ in the mass range $70<\mzams{}/\msun{}<150$, also varying the amount of mass loss through stellar winds to mimic the results expected from stars at lower metallicity. The fitting formula we implemented in \sevn{} gives results in agreement with those of W17. We obtain a comparable maximum BH mass (this work:$\sim 55\msun{}$, W17:\,$58\msun{}$), and a similar mass range where a dearth of compact remnants is observed (this work: $55\lesssim \mrem{}/\mzams{}\lesssim 120$, W17:\,$58<\mrem{}/\msun{}<133$). The differences between this work and W17 are due to the parameters $\mhef{}$ and $\mathcal{F}$ used for the PPISNe and PISNe fitting formula (see Appendix \ref{app:A}). Indeed, the values of $\mhef{}$ and $\mfin{}$, for different $\mzams{}$ and metallicity, strongly depend on the adopted stellar evolutionary models.

\section{Stellar evolution tables}
\label{app:D}

\begin{table}
	\caption{Final properties of stars of different initial mass and $Z=2.0\times 10^{-2}$ calculated with \sevn{}. \mzams{}: initial mass; \mfin{}: final mass; \mhef{}: final Helium core mass; \mcof{}: final Carbon--Oxygen core mass; \mrem{}(no PSNe): mass of the compact remnant when PPISNe and PISNe are switched off; \mrem{}: mass of the compact remnant. All values are given in \msun{}.}
	\label{tab:rem_zsun}
	\begin{tabular}{cccccc}
		\hline
		\mzams{} & \mfin{} & \mhef{} & \mcof{} & \mrem{} (no PSNe) & \mrem{} \\
		\hline
		8.0 & 8.0 & 2.2 & 1.2 & 1.3 & 1.3 \\
		13.8 & 13.0 & 4.5 & 2.6 & 1.3 & 1.3 \\
		19.5 & 17.1 & 6.9 & 4.4 & 3.4 & 3.4 \\
		25.2 & 19.9 & 9.8 & 6.4 & 8.1 & 8.1 \\
		31.0 & 16.0 & 12.6 & 8.8 & 10.7 & 10.7 \\
		36.8 & 13.1 & 13.1 & 9.5 & 9.8 & 9.8 \\
		42.5 & 13.4 & 13.4 & 9.6 & 10.2 & 10.2 \\
		48.2 & 14.7 & 14.7 & 10.6 & 12.6 & 12.6 \\
		54.0 & 16.3 & 16.3 & 11.7 & 14.6 & 14.6 \\
		59.8 & 18.1 & 18.1 & 13.1 & 16.3 & 16.3 \\
		65.5 & 20.2 & 20.2 & 14.5 & 18.2 & 18.2 \\
		71.2 & 22.6 & 22.6 & 16.2 & 20.3 & 20.3 \\
		77.0 & 23.9 & 23.9 & 17.2 & 21.5 & 21.5 \\
		82.8 & 25.2 & 25.2 & 18.1 & 22.7 & 22.7 \\
		88.5 & 26.6 & 26.6 & 19.1 & 23.9 & 23.9 \\
		94.2 & 27.4 & 27.4 & 19.7 & 24.7 & 24.7 \\
		100.0 & 27.7 & 27.7 & 19.8 & 25.0 & 25.0 \\
		105.8 & 26.1 & 26.1 & 18.7 & 23.5 & 23.5 \\
		111.5 & 24.5 & 24.5 & 17.6 & 22.1 & 22.1 \\
		117.2 & 23.0 & 23.0 & 16.5 & 20.7 & 20.7 \\
		123.0 & 21.7 & 21.7 & 15.6 & 19.5 & 19.5 \\
		128.8 & 20.4 & 20.4 & 14.7 & 18.4 & 18.4 \\
		134.5 & 19.3 & 19.3 & 13.9 & 17.4 & 17.4 \\
		140.2 & 18.4 & 18.4 & 13.2 & 16.5 & 16.5 \\
		146.0 & 17.5 & 17.6 & 12.7 & 15.8 & 15.8 \\
		151.8 & 16.7 & 16.9 & 12.2 & 15.0 & 15.0 \\
		157.5 & 16.5 & 16.5 & 11.9 & 14.8 & 14.8 \\
		163.2 & 16.4 & 16.4 & 11.9 & 14.8 & 14.8 \\
		169.0 & 16.4 & 16.4 & 11.8 & 14.7 & 14.7 \\
		174.8 & 16.3 & 16.3 & 11.8 & 14.7 & 14.7 \\
		180.5 & 16.3 & 16.3 & 11.7 & 14.7 & 14.7 \\
		186.3 & 16.2 & 16.2 & 11.7 & 14.6 & 14.6 \\
		192.0 & 16.2 & 16.2 & 11.7 & 14.6 & 14.6 \\
		197.8 & 16.2 & 16.2 & 11.7 & 14.5 & 14.5 \\
		203.5 & 16.1 & 16.1 & 11.6 & 14.5 & 14.5 \\
		209.2 & 16.1 & 16.1 & 11.6 & 14.5 & 14.5 \\
		215.0 & 16.2 & 16.2 & 11.7 & 14.6 & 14.6 \\
		220.8 & 16.2 & 16.2 & 11.7 & 14.6 & 14.6 \\
		226.5 & 16.2 & 16.2 & 11.7 & 14.6 & 14.6 \\
		232.2 & 16.2 & 16.2 & 11.7 & 14.6 & 14.6 \\
		238.0 & 16.2 & 16.2 & 11.7 & 14.6 & 14.6 \\
		243.8 & 16.3 & 16.3 & 11.7 & 14.6 & 14.6 \\
		249.5 & 16.3 & 16.3 & 11.7 & 14.7 & 14.7 \\
		255.2 & 16.4 & 16.4 & 11.8 & 14.7 & 14.7 \\
		261.0 & 16.5 & 16.5 & 11.9 & 14.8 & 14.8 \\
		266.8 & 16.6 & 16.6 & 11.9 & 14.9 & 14.9 \\
		272.5 & 16.7 & 16.7 & 12.0 & 15.0 & 15.0 \\
		278.2 & 16.8 & 16.8 & 12.1 & 15.1 & 15.1 \\
		284.0 & 16.9 & 16.9 & 12.2 & 15.2 & 15.2 \\
		289.8 & 17.0 & 17.0 & 12.2 & 15.3 & 15.3 \\
		295.5 & 17.0 & 17.0 & 12.3 & 15.3 & 15.3 \\
		301.2 & 17.0 & 17.0 & 12.3 & 15.3 & 15.3 \\
		307.0 & 17.0 & 17.0 & 12.3 & 15.3 & 15.3 \\
		312.8 & 17.0 & 17.0 & 12.3 & 15.3 & 15.3 \\
		318.5 & 17.0 & 17.0 & 12.3 & 15.3 & 15.3 \\
		324.2 & 17.0 & 17.0 & 12.2 & 15.3 & 15.3 \\
		330.0 & 17.0 & 17.0 & 12.2 & 15.3 & 15.3 \\
		335.8 & 16.9 & 16.9 & 12.2 & 15.2 & 15.2 \\
		341.5 & 16.9 & 16.9 & 12.2 & 15.2 & 15.2 \\
		350.0 & 16.9 & 16.9 & 12.2 & 15.2 & 15.2 \\
		\hline
	\end{tabular}
\end{table}

\begin{table}
	\caption{Same as Tab. \ref{tab:rem_zsun} but for $Z=2.0\times 10^{-3}$.}
	\label{tab:rem_01zsun}
	\begin{tabular}{cccccc}
		\hline
		\mzams{} & \mfin{} & \mhef{} & \mcof{} & \mrem{} (no PSNe) & \mrem{} \\
		\hline
		8.0 & 8.0 & 2.5 & 1.3 & 1.3 & 1.3 \\
		13.8 & 13.7 & 4.8 & 2.7 & 1.4 & 1.4 \\
		19.5 & 19.2 & 7.3 & 4.6 & 4.1 & 4.1 \\
		25.2 & 24.7 & 10.1 & 6.6 & 10.4 & 10.4 \\
		31.0 & 29.7 & 13.1 & 8.9 & 19.9 & 19.9 \\
		36.8 & 34.3 & 16.2 & 11.4 & 30.9 & 30.9 \\
		42.5 & 38.0 & 19.4 & 13.7 & 34.2 & 34.2 \\
		48.2 & 41.2 & 22.8 & 16.2 & 37.1 & 37.1 \\
		54.0 & 45.4 & 26.2 & 18.6 & 40.9 & 40.9 \\
		59.8 & 49.8 & 29.9 & 21.3 & 44.8 & 44.8 \\
		65.5 & 49.5 & 34.4 & 24.5 & 44.5 & 35.4 \\
		71.2 & 52.4 & 37.3 & 26.6 & 47.2 & 27.2 \\
		77.0 & 46.0 & 41.7 & 29.8 & 41.4 & 29.9 \\
		82.8 & 45.6 & 45.1 & 31.9 & 41.1 & 36.6 \\
		88.5 & 47.3 & 47.3 & 34.0 & 42.6 & 38.0 \\
		94.2 & 48.0 & 48.0 & 34.8 & 43.2 & 38.3 \\
		100.0 & 48.6 & 48.6 & 35.2 & 43.8 & 38.6 \\
		105.8 & 49.7 & 49.7 & 35.9 & 44.8 & 39.1 \\
		111.5 & 50.8 & 50.8 & 36.7 & 45.7 & 39.6 \\
		117.2 & 51.8 & 51.8 & 37.4 & 46.6 & 40.0 \\
		123.0 & 53.6 & 53.6 & 38.9 & 48.2 & 40.8 \\
		128.8 & 56.1 & 56.1 & 41.0 & 50.5 & 41.4 \\
		134.5 & 58.5 & 58.5 & 43.0 & 52.7 & 30.0 \\
		140.2 & 60.8 & 60.8 & 45.0 & 54.7 & 18.1 \\
		146.0 & 63.0 & 63.0 & 46.9 & 56.7 & 5.7 \\
		151.8 & 65.2 & 65.2 & 48.7 & 58.7 & 0.0 \\
		157.5 & 67.4 & 67.4 & 50.2 & 60.6 & 0.0 \\
		163.2 & 69.4 & 69.4 & 51.6 & 62.5 & 0.0 \\
		169.0 & 71.5 & 71.5 & 53.0 & 64.3 & 0.0 \\
		174.8 & 73.4 & 73.4 & 54.3 & 66.1 & 0.0 \\
		180.5 & 75.3 & 75.3 & 55.5 & 67.8 & 0.0 \\
		186.3 & 77.1 & 77.1 & 56.8 & 69.4 & 0.0 \\
		192.0 & 78.9 & 78.9 & 57.9 & 71.0 & 0.0 \\
		197.8 & 80.7 & 80.7 & 59.2 & 72.6 & 0.0 \\
		203.5 & 82.6 & 82.6 & 60.8 & 74.4 & 0.0 \\
		209.2 & 84.9 & 84.9 & 62.9 & 76.4 & 0.0 \\
		215.0 & 87.1 & 87.1 & 64.9 & 78.4 & 0.0 \\
		220.8 & 89.2 & 89.2 & 66.9 & 80.3 & 0.0 \\
		226.5 & 91.3 & 91.3 & 68.9 & 82.2 & 0.0 \\
		232.2 & 93.3 & 93.3 & 70.8 & 84.0 & 0.0 \\
		238.0 & 95.3 & 95.3 & 72.7 & 85.8 & 0.0 \\
		243.8 & 97.2 & 97.2 & 74.6 & 87.5 & 0.0 \\
		249.5 & 99.2 & 99.2 & 76.4 & 89.3 & 0.0 \\
		255.2 & 101.3 & 101.3 & 78.1 & 91.2 & 0.0 \\
		261.0 & 103.5 & 103.5 & 79.6 & 93.1 & 0.0 \\
		266.8 & 105.6 & 105.6 & 81.1 & 95.1 & 0.0 \\
		272.5 & 107.7 & 107.7 & 82.5 & 97.0 & 0.0 \\
		278.2 & 109.8 & 109.8 & 83.9 & 98.8 & 0.0 \\
		284.0 & 111.8 & 111.8 & 85.3 & 100.6 & 0.0 \\
		289.8 & 113.8 & 113.8 & 86.7 & 102.4 & 0.0 \\
		295.5 & 115.7 & 115.7 & 88.0 & 104.1 & 0.0 \\
		301.2 & 117.7 & 117.7 & 89.6 & 105.9 & 0.0 \\
		307.0 & 119.7 & 119.7 & 91.4 & 107.7 & 0.0 \\
		312.8 & 121.7 & 121.7 & 93.3 & 109.5 & 0.0 \\
		318.5 & 123.6 & 123.6 & 95.2 & 111.3 & 0.0 \\
		324.2 & 125.6 & 125.6 & 97.0 & 113.0 & 0.0 \\
		330.0 & 127.5 & 127.5 & 98.9 & 114.7 & 0.0 \\
		335.8 & 129.3 & 129.3 & 100.7 & 116.4 & 0.0 \\
		341.5 & 131.1 & 131.1 & 102.5 & 118.0 & 0.0 \\
		350.0 & 132.4 & 132.4 & 103.3 & 119.2 & 0.0 \\
		\hline
	\end{tabular}
\end{table}

\begin{table}
	\caption{Same as Tab. \ref{tab:rem_zsun} but for $Z=2.0\times 10^{-4}$.}
	\label{tab:rem_001zsun}
	\begin{tabular}{cccccc}
		\hline
		\mzams{} & \mfin{} & \mhef{} & \mcof{} & \mrem{} (no PSNe) & \mrem{} \\
		\hline
		8.0 & 8.0 & 2.5 & 1.4 & 1.3 & 1.3 \\
		13.8 & 13.7 & 4.8 & 2.8 & 1.4 & 1.4 \\
		19.5 & 19.5 & 7.2 & 4.5 & 4.0 & 4.0 \\
		25.2 & 25.2 & 10.1 & 6.7 & 10.7 & 10.7 \\
		31.0 & 31.0 & 13.2 & 9.1 & 21.3 & 21.3 \\
		36.8 & 36.7 & 16.4 & 11.5 & 33.0 & 33.0 \\
		42.5 & 42.4 & 19.7 & 14.0 & 38.1 & 38.1 \\
		48.2 & 48.0 & 23.2 & 16.6 & 43.2 & 43.2 \\
		54.0 & 53.7 & 26.8 & 19.2 & 48.3 & 48.3 \\
		59.8 & 59.3 & 30.3 & 21.8 & 53.4 & 53.4 \\
		65.5 & 64.6 & 33.5 & 24.2 & 58.2 & 48.5 \\
		71.2 & 70.1 & 37.0 & 26.7 & 63.1 & 28.8 \\
		77.0 & 75.3 & 40.3 & 29.2 & 67.7 & 31.1 \\
		82.8 & 79.5 & 44.1 & 32.0 & 71.5 & 33.8 \\
		88.5 & 83.1 & 48.3 & 35.2 & 74.8 & 36.6 \\
		94.2 & 88.1 & 51.9 & 37.8 & 79.3 & 39.2 \\
		100.0 & 94.0 & 55.5 & 40.3 & 84.6 & 41.9 \\
		105.8 & 100.0 & 59.3 & 43.5 & 90.0 & 44.8 \\
		111.5 & 105.7 & 63.0 & 46.8 & 95.2 & 12.5 \\
		117.2 & 111.2 & 66.5 & 49.8 & 100.1 & 0.0 \\
		123.0 & 116.9 & 70.1 & 52.8 & 105.2 & 0.0 \\
		128.8 & 122.8 & 73.8 & 55.6 & 110.5 & 0.0 \\
		134.5 & 128.5 & 77.3 & 58.3 & 115.6 & 0.0 \\
		140.2 & 133.9 & 80.8 & 60.8 & 120.5 & 0.0 \\
		146.0 & 139.0 & 84.1 & 63.3 & 125.1 & 0.0 \\
		151.8 & 144.5 & 87.6 & 66.0 & 130.1 & 0.0 \\
		157.5 & 150.5 & 91.3 & 69.1 & 135.4 & 0.0 \\
		163.2 & 156.3 & 95.1 & 72.1 & 140.7 & 0.0 \\
		169.0 & 161.9 & 98.7 & 75.1 & 145.8 & 0.0 \\
		174.8 & 167.4 & 102.2 & 77.9 & 150.7 & 0.0 \\
		180.5 & 172.7 & 105.6 & 80.7 & 155.4 & 0.0 \\
		186.3 & 177.8 & 108.9 & 83.5 & 160.0 & 0.0 \\
		192.0 & 182.7 & 112.1 & 86.1 & 164.5 & 0.0 \\
		197.8 & 187.7 & 115.4 & 88.9 & 168.9 & 0.0 \\
		203.5 & 192.9 & 118.8 & 91.8 & 173.6 & 0.0 \\
		209.2 & 198.4 & 122.5 & 95.1 & 178.6 & 0.0 \\
		215.0 & 203.9 & 126.1 & 98.3 & 183.5 & 0.0 \\
		220.8 & 209.2 & 129.7 & 101.5 & 188.2 & 0.0 \\
		226.5 & 214.3 & 133.2 & 104.6 & 192.9 & 0.0 \\
		232.2 & 219.3 & 136.5 & 107.7 & 197.4 & 197.4 \\
		238.0 & 224.2 & 139.9 & 110.7 & 201.8 & 201.8 \\
		243.8 & 229.0 & 143.2 & 113.7 & 206.1 & 206.1 \\
		249.5 & 233.8 & 146.4 & 116.6 & 210.4 & 210.4 \\
		255.2 & 238.9 & 150.1 & 119.3 & 215.0 & 215.0 \\
		261.0 & 244.1 & 153.7 & 122.0 & 219.7 & 219.7 \\
		266.8 & 249.1 & 157.3 & 124.6 & 224.2 & 224.2 \\
		272.5 & 254.1 & 160.9 & 127.1 & 228.7 & 228.7 \\
		278.2 & 259.0 & 164.3 & 129.6 & 233.1 & 233.1 \\
		284.0 & 263.7 & 167.8 & 132.0 & 237.3 & 237.3 \\
		289.8 & 268.4 & 171.1 & 134.4 & 241.5 & 241.5 \\
		295.5 & 272.9 & 174.5 & 136.8 & 245.6 & 245.6 \\
		301.2 & 277.7 & 178.0 & 139.7 & 249.9 & 249.9 \\
		307.0 & 282.3 & 181.5 & 142.7 & 254.1 & 254.1 \\
		312.8 & 287.0 & 185.2 & 146.0 & 258.3 & 258.3 \\
		318.5 & 291.6 & 188.7 & 149.3 & 262.4 & 262.4 \\
		324.2 & 296.1 & 192.2 & 152.6 & 266.5 & 266.5 \\
		330.0 & 300.6 & 195.7 & 155.8 & 270.5 & 270.5 \\
		335.8 & 304.9 & 199.1 & 159.0 & 274.4 & 274.4 \\
		341.5 & 309.3 & 202.1 & 162.2 & 278.3 & 278.3 \\
		350.0 & 311.6 & 204.4 & 165.4 & 281.2 & 281.2 \\
		\hline
	\end{tabular}
\end{table}

Tables \ref{tab:rem_zsun}, \ref{tab:rem_01zsun}, and \ref{tab:rem_001zsun} show the final parameters of various progenitor stars and the mass of their compact remnants when PPISNe and PISNe are turned on and when PPISNe and PISNe are switched off in \sevn{}, for $Z=2.0\times 10^{-2}$, $Z=2.0\times 10^{-3}$, and $Z=2.0\times 10^{-4}$, respectively. The adopted SN explosion mechanism is the delayed model and the stellar evolution tracks come from the PARSEC code.

	\bsp  
	\label{lastpage}
\end{document}